\documentstyle[a4,psfig]{article}
\begin{document}

\title{On universality in ageing ferromagnets}

\author{Christophe Chatelain \\
Laboratoire de Physique des Mat\'eriaux,\\
Universit\'e Henri Poincar\'e Nancy I,\\
BP~239, Boulevard des aiguillettes,\\
F-54506 Vand{\oe}uvre l\`es Nancy Cedex, France\\
e-mail: chatelai@lpm.u-nancy.fr}

\maketitle

\begin{abstract}
This work is a contribution to the study of universality in
out-of-equilibrium lattice models undergoing a second-order phase transition
at equilibrium. The experimental protocol that we have chosen is the following:
the system is prepared in its high-temperature phase and then quenched at the
critical temperature $T_c$. We investigated by mean of Monte Carlo
simulations two quantities that are believed to take universal values: the
exponent $\lambda/z$ obtained from the decay
of autocorrelation functions and the asymptotic value $X_\infty$ of the
fluctuation-dissipation ratio $X(t,s)$. This protocol was applied to the Ising
model, the 3-state clock model and the 4-state Potts model on square, triangular
and honeycomb lattices and to the Ashkin-Teller model at the point belonging
at equilibrium to the 3-state Potts model universality class and to
a multispin Ising model and the Baxter-Wu model both belonging to the 4-state
Potts model universality class at equilibrium.
\end{abstract}


\def\build#1_#2^#3{\mathrel{
\mathop{\kern 0pt#1}\limits_{#2}^{#3}}}
\def\ket#1{\left| #1 \right\rangle}
\def\bra#1{\left\langle #1\right|}
\def\braket#1#2{\left\langle\vphantom{#1#2} #1\right.
\left| \vphantom{#1#2}#2\right\rangle}
\def\spring{\hskip 0pt minus 1fil}
\def\identite{{\rm 1}\hbox to 1 pt{\spring\rm l}}

\section{Introduction}
Universality is an extremely fruitful concept in statistical physics and
has been widely studied in the context of systems undergoing a second order
phase transition. At thermodynamic equilibrium, the length scale $\xi$ of
spatial correlation functions of the local order parameter diverges as a
power-law with a critical exponent $\nu$ as the temperature approaches the
critical temperature. Other observables (energy, magnetisation,~$\ldots$)
display such a power-law dependence as well but with different critical
exponents. It turns out that many microscopic details of the Hamiltonian
do not change the value of these critical exponents.
The big success of the renormalisation group has been to explain that a few
number of these critical exponents are independent and that different models
have the same set of critical exponents if they differ only by irrelevant
operators. Consequently, the usual models of statistical physics can be
classified into universality classes according to the value of their critical
exponents. The space and order parameter dimensions, the Hamiltonian symmetries,
the presence of long-range interactions, of randomness~$\ldots$ determine the
universality class. Not only critical exponents but also ratios of critical
amplitudes turn out to be universal.

The question of universality in out-of-equilibrium processes has been
addressed in the context of dynamical transitions undergone by reaction-diffusion
systems~\cite{Odor02}. A set of exponents can be defined, for instance
using the algebraic decay with time of the density of active sites.

In the following, we will restrict ourselves to another important set of
out-of-equilibrium processes provided by systems that undergo a second-order
phase transition at equilibrium and that are quenched from their
high-temperature phase to their critical point or their low-temperature phase.
Because of the competition of domains in different equilibrium states,
such systems never reach equilibrium in the thermodynamic limit~\cite{Bray94}.
The hypothesis was made that systems belonging to the same universality
class at equilibrium share universal quantities if their time-evolution
is governed by dynamics satisfying the same conservation laws and if
order parameter and conserved quantities are related in the
same way~\cite{Hohenberg77}. In the following, we will consider systems
belonging to the same universality class at equilibrium and whose dynamics
has no conserved quantity (model A). In this context new exponents have been
defined: the dynamical exponent $z$ related to the growth of the domain
length scale $L(t)$ with time by~\cite{Hohenberg77} 
	\begin{equation}
		L(t)\sim t^{1/z}
	\label{Eq_Def_z}
	\end{equation}
and the autocorrelation exponent $\lambda$ related to the decay of two-time
autocorrelation functions of the local order parameter~\cite{Fisher88, Huse89} by
	\begin{equation}
		C(t,s)\sim t^{-\lambda/z}.
	\label{Eq_Def_Lambda}
	\end{equation}
These two new exponents $z$ and $\lambda$ take different values whether the
system is quenched at its critical temperature $T_c$ or below. They
are believed to be universal. Numerical calculations for the Ising model on
square, triangular and honeycomb lattices~\cite{Wang97} or for three different
models belonging to the Ising equilibrium universality class~\cite{Nightingale00}
support this conjecture for the exponent $z$ at $T=T_c$. Simulations for other
models, for instance the 3-state Potts model~\cite{Schuelke95}, give estimate
for $z$ sufficiently close to give support to the conjecture of a dynamic
universality.

The decay of persistence, i.e. the probability that the total order parameter
has not changed sign at time $t$, defines a third new non-trivial exponent
$\theta'$. In the rather unusual case where the dynamics of the total order
parameter is Markovian, it is related to the two previously-defined ones by
$z\theta'=\lambda-d+1-\eta/2$~\cite{Majumdar96}. 

Renormalisation-group study of the $O(n)$ model in $d=4-\epsilon$
dimension shows that if the system is quenched at the critical temperature
from an initial state with a small non-vanishing magnetisation, the latter
first grows as $t^\theta$ before decaying asymptotically as
$\xi^{-\beta/\nu}\sim L(t)^{-\beta/\nu}\sim t^{-\beta/\nu z}$~\cite{Janssen89}.
The initial-slip critical exponent $\theta$ is related to the autocorrelation
exponent by
	\begin{equation}
		\lambda=d-z\theta
	\end{equation}
where $d$ is the dimension of the space. This behaviour of the magnetisation has
been exploited in the so-called short-time dynamics Monte Carlo
method~\cite{Li95}. The question of universality has been addressed by several
authors in this context. The Ashkin-Teller model has been studied by short-time
dynamics for several points of its exactly-known critical line~\cite{Li97}.
Unfortunately, the point belonging to the 3-state Potts model universality class,
i.e. $y=3/4$ where $\nu={2-y\over 3-2y}$, has not been considered. Noting that the
estimates of $\theta$ given by the authors vary roughly linearly with the parameter
$y$, one can estimate the exponent $\theta$ at the point $y=3/4$ to be $\theta\simeq 0.111$.
This result is not compatible with that obtained for the 3-state Potts
model~\cite{Schuelke95}: $\theta\simeq 0.0815(27)$. Moreover, numerical
estimates of this exponent $\theta$ for the Baxter-Wu~\cite{Arashiro03} and a multispin
Ising model~\cite{Simoes01} both belonging to the 4-state Potts model equilibrium
universality class are incompatible with estimates for the latter while the dynamical
exponent $z$ seems to be the same~\cite{Arashiro03}.

Apart from exponents, ratios of amplitudes have also been conjectured
to be universal, as for example the fluctuation-dissipation ratio $X(t,s)$
in the asymptotic limit $t,s\rightarrow +\infty$.
The response function $R(t,s)$ to an infinitesimal field $h(s)$ coupled to the
local order parameter is expected to decay algebraically with time with the
same exponent $\lambda/z$ as autocorrelation functions provided the initial state
does not display spatial correlations. Note that autocorrelation and response
functions are related at equilibrium by the fluctuation-dissipation theorem (FDT):
	\begin{equation}
		k_BT\ \! R(t,s)={\partial\over\partial s}C(t,s).
	\label{FDT}
	\end{equation}
On the basis of a mean-field study of a spin-glass-like model, it was conjectured
that the fluctuation-dissipation theorem may be generalised by adding to
Eq.~\ref{FDT} a multiplicative factor depending on time only through autocorrelation
functions~\cite{Cugliandolo94}:
	\begin{equation}
		k_BT\ \! R(t,s)\sim  X\big[C(t,s)\big]
		{\partial\over\partial s}C(t,s),\quad (t\sim s\gg 1).
	\label{gFDT}
	\end{equation}
However, numerical simulations for the Ising model~\cite{Chatelain03} and
renormalisation-group calculations for the $O(n)$ model~\cite{Gambassi02}
suggest that the fluctuation-dissipation ratio is not a function of $C(t,s)$ only
but of $t/s$. Scaling arguments constrain the autocorrelation and response
function to the following asymptotic behaviour~\cite{Godreche00b} at $T_c$:
	\begin{equation}
		C(t,s)\sim s^{-{2\beta\over\nu z}}f_C\left({t\over s}\right),
		\quad
		k_BT_c\ \! R(t,s)\sim s^{-1-{2\beta\over\nu z}}
		f_R\left({t\over s}\right)
		\quad (t\sim s\gg 1)
	\label{ScalingCR}
	\end{equation}
where both $f_C(x)$ and $f_R(x)$ are scaling functions whose asymptotic behaviour
is given by
	\begin{equation}
		f_{C/R}(x)\sim A_{C/R}\ \!x^{-\lambda/z},
		\quad x\rightarrow +\infty.
	\label{ScalingCR2}
	\end{equation}
Inserting Eq. \ref{ScalingCR} and \ref{ScalingCR2} into Eq. \ref{gFDT} leads to the
asymptotic value of the fluctuation-dissipation ratio:
	\begin{equation}
		X_{\infty}=\lim_{t,s\rightarrow +\infty} X(t,s)
		={A_R\over A_C}\left[{\lambda\over z}-{2\beta\over\nu z}\right]^{-1}
	\end{equation}
which turns out to depend only on exponents which are believed to be universal
and on the ratio of autocorrelation and response amplitudes. The latter has been
conjectured to be universal~\cite{Godreche00b}. This conjecture applies thus
to $X_{\infty}$ as well. Renormalisation-group calculations~\cite{Gambassi02} of
the $O(1)$ model gives indeed an estimate of $X_\infty$ compatible with
numerical values for the Ising model~\cite{Berthier03, Chatelain03, Sastre03}.
Moreover, a recent numerical calculation of the integrated response
function to an exchange coupling perturbation gave an estimate of $X_{\infty}$ in
full agreement with theses values~\cite{Berthier03}. The Ising model may
be peculiar since a one-loop renormalisation-group calculation of the $O(1)$ model
gives the same value of $X_{\infty}$ whether the order parameter is coupled to a
conserved quantity (model C) or not (model A)~\cite{Gambassi03}. Let us mention that
calculations for the Ising-Glauber chain indicates that $X_{\infty}$ takes a
non-vanishing value only with a well defined protocol~: a quench at the critical
temperature from an initial state with an infinite number of domain
walls, i.e. from an initial disordered state~\cite{Henkel04}.
\\

The plan of this paper is the following: in the first section, the expressions
of the Hamiltonians of the different models we studied are given. The Glauber
dynamics is then defined and we review the method we used to calculate
the fluctuation-dissipation ratio without resorting to the Cugliandolo
conjecture (Eq~\ref{gFDT}). The second section is devoted to the
characterisation of the three universality classes under consideration.
The question of the influence of the lattice on the quantities
that are supposed to be universal is addressed in the third section.
The fourth section is devoted to the comparison of these universal quantities
for different models belonging to the same universality class at equilibrium.

\section{Dynamics and Models}
\subsection{Definition of the models}
We have considered two-dimensional lattice models belonging at equilibrium to
three different universality classes. The first of them is the
Ising model~\cite{Ising25} defined by the Hamiltonian:
	\begin{equation}{\cal H}_{\rm Ising}
	=-J\sum_{(i,j)} \sigma_i\sigma_j-\sum_i h_i\sigma_i,
	\quad \sigma_i=\pm 1\end{equation}
where the first sum extends to nearest neighbours on the lattice. A local magnetic
field coupled to the local order parameter $m_i=\sigma_i$ has been added to allow
for the definition of a response function. For vanishing magnetic field, the model
undergoes a second-order phase transition associated to the breaking of a
$Z(2)$-symmetry. Duality relation leads to the critical point on the square lattice
$\beta_c J={1\over 2}\ln(1+\sqrt 2)$ where $\beta_c=1/k_BT_c$~\cite{Kramers41}. Although
triangular and honeycomb lattices are not self-dual but dual of each other, the exact
determination of the critical temperature is nevertheless possible~\cite{Kim74}
and gives $\beta_c J\simeq 0.2746$ on a triangular lattice and
$\beta_c J\simeq 0.6585$ on the honeycomb lattice.

The second universality class we have considered is that of the 3-state Potts
model~\cite{Potts52} whose Hamiltonian is
	\begin{equation}
	{\cal H}_{\rm Potts}
	=-J\sum_{(i,j)} \delta_{\sigma_i,\sigma_j}-\sum_i h_i\delta_{\sigma_i,0},
	\quad \sigma_i=0,\ldots q-1
	\end{equation}
where the number of state $q$ is set to $q=3$. For vanishing magnetic field, the model
undergoes a second-order phase transition associated to the breaking of a
$Z(3)$-symmetry. Self-duality relation allows for the exact determination of the critical
point: $\beta_c J=\ln(1+\sqrt q)$ at zero-magnetic field on the square
lattice. Critical points can also be obtained for the triangular and honeycomb
lattices~\cite{Kim74}. In order to achieve better numerical stability, we used an
equivalent formulation of this model, known as the clock model, whose order parameter
is vanishing at the critical temperature without resorting to any additional normalisation:
	\begin{eqnarray}
	{\cal H}_{\rm Clock}
	=&-&J\sum_{(i,j)} \cos\left({2\pi\over q}(\sigma_i-\sigma_j)\right)
	\nonumber\\
	&-&\sum_i \left[h_i^x\cos\left({2\pi\over q}\sigma_i\right)
	+h_i^y\sin\left({2\pi\over q}\sigma_i\right)\right]
	\quad\sigma_i=0,\ldots q-1.
	\end{eqnarray}
The critical temperature is readily obtained to be two-third of that of the
equivalent $q$-state Potts model.
As a prototype of model belonging to this universality class, we have
chosen the Ashkin-Teller model~\cite{Ashkin43} defined by the Hamiltonian
	\begin{equation}
	{\cal H}_{\rm AT}=-J\sum_{(i,j)}\sigma_i\sigma_j
	-J'\sum_{(i,j)} \tau_i\tau_j-K\sum_{(i,j)}\sigma_i\sigma_j\tau_i\tau_j
	-\sum_i h_i\sigma_i,\quad \sigma_i,\tau_i=\pm 1.
	\end{equation}
Indeed, in the isotropic case $J=J'$ the universality class changes from that
of the Ising model to that of the 4-state Potts model along an exactly-known
critical line. The point belonging to the 3-state Potts model is defined by the
critical couplings: $\beta_c J=\beta_c J'\simeq 0.34763$ and
$\beta_c K\simeq 0.14209$~\cite{Fan72}.

The third universality class we have considered is that of the 4-state Potts
model. For vanishing magnetic field, the model undergoes a second-order phase
transition associated to the breaking of a $Z(4)$-symmetry. Duality relations allows
for the determination of the critical point on the square, triangular and
honeycomb lattices~\cite{Kim74}.
Apart from the 4-state Potts model, we have made calculations for two
other models belonging to this universality class: the Baxter-Wu model~\cite{Baxter73}
defined by the Hamiltonian
	\begin{equation}
	{\cal H}_{\rm BW}=-J\sum_{(i,j,k)} \sigma_i\sigma_j\sigma_k
	-\sum_i h_i\sigma_i,\quad \sigma_i=\pm 1
	\end{equation}
where the sum extend over all triangles of a triangular lattice and
a multispin Ising model~\cite{Turban82} defined on a square lattice by the Hamiltonian
	\begin{equation}
	{\cal H}_{\rm DT}=-J\sum_{(i,j)}' \sigma_i\sigma_j
	-K\sum_{(i,j,k)}'' \sigma_i\sigma_j\sigma_k-\sum_i h_i\sigma_i,
	\quad \sigma_i=\pm 1
	\end{equation}
where the first sum extends over couples of neighbour spins along the horizontal
direction and the second over triplets of spins aligned along the
vertical direction. For the Baxter-Wu model, the ground states correspond to
spin configurations where triangle vertices are decorated with spins
$\{+,+,+\}$, $\{+,-,-\}$, $\{-,+,-\}$ or $\{-,-,+\}$. In the case of the multispin Ising
model, the ground-states correspond to rows decorated with the sequence of spins:
$\{+,+,+\}$, $\{+,-,-\}$, $\{-,+,-\}$ or $\{-,-,+\}$. Duality relations lead for
the Baxter-Wu model to the critical point $\beta_c J\simeq {1\over 2}\ln(1+\sqrt 2)$ 
and for the multispin Ising model to the same critical line in the $J-K$ plane
than the anisotropic Ising model~\cite{Turban82} from which we have chosen the point
$\beta_c J=1$, $\beta_c K\simeq 0.1362$.
 
\subsection{Glauber dynamics and observables}
The time evolution of all these models is governed by the same
discrete-time Glauber dynamics. Let us denote by $\wp(\{\sigma\},t)$ the
probability of the spin configuration $\{\sigma\}$ at time $t$. The dynamics
is defined by the usual master equation for Monte Carlo simulations:
	\begin{equation}
        \wp(\{\sigma\},t+1)=\sum_{\{\sigma'\}}
	W(\{\sigma'\}\rightarrow \{\sigma\},t)\wp(\{\sigma'\},t)
  	\label{eq1}
\end{equation}
where $W(\{\sigma\}\rightarrow \{\sigma'\},t)$ is the transition rate per
time step from the state $\{\sigma\}$ to the state $\{\sigma'\}$ at time $t$.
The conditional probability $\wp(\{\sigma\},t|\{\sigma'\},s)$ for the
system to be in the state $\{\sigma\}$ at time $t$ knowing that it was in the
state $\{\sigma'\}$ at time $s$ satisfies the master equation too.
The condition of stationarity ${\partial\over\partial t}\wp_{\rm eq}(\{\sigma\})=0$
of the equilibrium Boltzmann probability $\wp_{\rm eq}(\{\sigma\})$ is ensured by
the so-called detailed balance condition
\begin{equation}
  	\wp_{\rm eq}(\{\sigma'\})W(\{\sigma'\}\rightarrow \{\sigma\},t)
  	=\wp_{\rm eq}(\{\sigma\})W(\{\sigma\}\rightarrow \{\sigma'\},t).
  	\label{eq2}
\end{equation}
This last unnecessary but sufficient condition is fulfilled by
the single-spin flip dynamics defined for the single-spin flip
$\sigma_i\rightarrow\tilde\sigma_i\ne \sigma_i$, where
$\tilde\sigma_i$ is a randomly chosen spin, whose transition rates are
\begin{equation}
  	W(\{\sigma\}\rightarrow \{\sigma'\},t)
	=\Big[\prod_{j\ne i}\delta_{\sigma_j,\sigma_j'}\Big]
	{\delta_{\sigma_i',\sigma_i}+\delta_{\sigma_i',\tilde\sigma_i}
	e^{-\beta\Delta E}\over 1+e^{-\beta\Delta E}}
  	\label{eq3}
\end{equation}
where $\tilde\sigma_i$ is a trial state for the spin $\sigma_i$ and
$\Delta E={\cal H}(\sigma_1,\ldots \tilde\sigma_i,\ldots\sigma_N)
-{\cal H}(\sigma_1,\ldots\sigma_i,\ldots\sigma_N)$ is the energy difference when
replacing $\sigma_i$ by $\tilde\sigma_i$. The Markov chain has to be averaged over
all possible spin-flips to recover Glauber dynamics~\cite{Glauber63}. This is
the usual way Monte Carlo simulations proceed. In the simulations to be presented
in the following, the sites on which spin flips are applied are randomly chosen.
Let us mention that another definition of the transition rates may be given
\begin{equation}
  	W(\{\sigma\}\rightarrow \{\sigma'\},t)
	=\Big[\prod_{j\ne i}\delta_{\sigma_j,\sigma_j'}\Big]
	{e^{-\beta{\cal H}(\{\sigma'\})}\over \sum_{\tilde\sigma_i}
	e^{-\beta{\cal H}(\sigma_1,\ldots \tilde\sigma_i,\ldots\sigma_N)}}
  	\label{eq3b}
\end{equation}
which corresponds to the Heat-Bath algorithm in Monte Carlo simulations.
\\

Autocorrelation functions of the local order parameter $m_i$ have been defined
as usual as
\begin{eqnarray}
	C_{ij}(t,s)&=&\langle m_i(t)m_j(s)\rangle				\\
	&=&\sum_{\{\sigma\},\{\sigma'\}} m_i(\{\sigma\})
	\wp(\{\sigma\},t|\{\sigma'\},s) m_j(\{\sigma'\})\wp(\{\sigma'\},s)	
	\nonumber
  	\label{eq4}
\end{eqnarray}
Since periodic boundary conditions are used, autocorrelation functions are
expected to be invariant under space translations. In the following, we will consider
averaged autocorrelation functions over the lattice:
$C(t,s)={1\over N}\sum_{i=1}^N  C_{ii}(t,s)$.
The response to an infinitesimal field $h_i$ coupled to the local order parameter
$m_i$ has been computed using a recently proposed method for the Ising-Glauber
model~\cite{Chatelain03}. We shall derive this result for a more general spin
lattice model. The coupling of the infinitesimal field to the local order
parameter is performed by modifying the transition rate Eq.~\ref{eq3} in the
following way:
\begin{equation}
  	 W(\{\sigma\}\rightarrow \{\sigma'\},t)
	=\Big[\prod_{j\ne i}\delta_{\sigma_j,\sigma_j'}\Big]
	{\delta_{\sigma_i',\sigma_i}e^{\beta h_im_i(\{\sigma\})}
	+\delta_{\sigma_i',\tilde\sigma_i}e^{-\beta\Delta E+\beta h_i
	m_i(\{\tilde\sigma\})}\over e^{\beta h_im_i(\{\sigma\})}
	+e^{-\beta\Delta E+\beta h_im_i(\{\tilde\sigma\})}}.
  	\label{eq5}
\end{equation}
This modification corresponds to the addition of the Zeeman Hamiltonian
$-h_im_i$ to the Boltzmann weight of the equilibrium probability distribution
in Eq. \ref{eq2}. The average order parameter at time $t$ can be expanded as:
\begin{eqnarray}
    \langle m_j(t)\rangle
    &=&\sum_{\{\sigma\}} m_j(\{\sigma\})\ \!\wp(\{\sigma\},t)		\\        
    &=&\sum_{\{\sigma\},\{\sigma'\}} m_j(\{\sigma\})\ \!
    \wp(\{\sigma\},t|\{\sigma'\},s+1)\wp(\{\sigma'\},s+1) \nonumber\\
    &=&\sum_{\{\sigma\},\{\sigma'\},\atop\{\sigma''\}} m_j(\{\sigma\})\ \!
	\wp(\{\sigma\},t|\{\sigma'\},s+1)W(\{\sigma''\}\rightarrow\{\sigma'\},s)
	\wp(\{\sigma''\},s).						\nonumber
  	\label{eq6}
\end{eqnarray}
The magnetic field being branched only during the time step $s$, the transition
rate $W(\{\sigma''\}\rightarrow\{\sigma'\},s)$ is the only quantity depending on
$h_i$ provided that the spin-flip occurring at time $s$ affects the spin branched
to the magnetic field. The derivative of the transition rate being
\begin{eqnarray}
 	&&\left(\partial W\over\partial h_i(s)\right)_{h=0}\!\!\!\!
	(\{\sigma''\}\rightarrow\{\sigma'\},s)			\\
	&&\hskip 2truecm
	=\beta\left[m_i(\{\sigma'\})-{m_i(\{\sigma\})+m_i(\{\tilde\sigma\})
	e^{-\beta\Delta E}\over 1+e^{-\beta\Delta E}}\right]\epsilon_i(s)
	W(\{\sigma''\}\rightarrow\{\sigma'\},s)			\nonumber
  	\label{eq7}
\end{eqnarray}
where $\epsilon_i(s)$ is equal to 1 if the transition rate involves a spin-flip on the
spin $\sigma_i$ at time $s$ and 0 otherwise, the response to the perturbation follows
\begin{eqnarray}
	 R_{ji}(t,s)
	=\left(\partial \langle m_j(t)\rangle\over\partial h_i(s)\right)_{h=0}
	&=&\sum_{\{\sigma\},\{\sigma'\}} m_j(\{\sigma\})\ \!
	\wp(\{\sigma\},t|\{\sigma'\},s+1)			\nonumber\\
	 &&\hskip 1truecm\times\left(\partial W\over\partial h_i(s)
	\right)_{h=0}(\{\sigma''\}\rightarrow\{\sigma'\},s)\epsilon_i(s)
	\wp(\{\sigma''\},s)					\nonumber\\
	 &=&\beta\langle m_j(t)\left[m_i(s+1)-m_i^{\rm W}(s)\right]\epsilon_i(s)\rangle
  	\label{eq8}
\end{eqnarray}
where $m_i^{\rm W}(s)$ is the average local order parameter over
$\sigma_i(s)$ and the trial value $\tilde\sigma_i$ that appeared in Eq.~\ref{eq6}.
The response is zero if the transition rate at time $s$ involved any other spin than
$\sigma_i$. This procedure is easily implemented in Monte Carlo simulations.
Since periodic boundary conditions are to be used, we will consider only averaged
response functions over the lattice: $R(t,s)={1\over N}\sum_{i=1}^N  R_{ii}(t,s)$.
Using the discrete-time master equation Eq~\ref{eq1} and the Boltzmann equilibrium
probability distribution, the FDT can be shown to be $k_BTR(t,s)=C(t,s+1)
-C(t,s)$ at equilibrium for the above-define discrete-time process~\cite{Chatelain03}.
The ratio measuring the violation of the FDT out-of-equilibrium is readily obtained to be:
\begin{equation}
	 X(t,s)={k_BTR(t,s)\over C(t,s+1)-C(t,s)}
	={\sum_i \langle m_i(t)\left[m_i(s+1)-m_i^{\rm W}(s)\right]\epsilon_i(s)\rangle
	\over \sum_i \langle m_i(t)\left[m_i(s+1)-m_i(s)\right]\rangle}.
  	\label{eq9}
\end{equation}
This method allows for the calculation of $X(t,s)$ for all values of both $s$ and $t$
during the same Monte Carlo simulation without resorting to Cugliandolo
conjecture Eq. \ref{gFDT} and without applying any finite magnetic field that could
induce non-linear response. This method has been applied to the calculation of the
integrated response function~\cite{Ricci03}, the study of the XY-model~\cite{Abriet03}
and of the Ising model with Kawasaki dynamics~\cite{Godreche04}. It turned out to
require smaller values of $s$ and $t$ than measurements of the integrated response
function to give an accurate estimation of $X_\infty$ but on the other hand, an
average over a larger number of histories is needed to give a stable estimate of
$X_\infty$.

\section{Ising and Potts universality classes}
We first present the study of the Ising model, the 3-state clock model and the
4-state Potts model which belong to different universality classes at equilibrium.
The system is first prepared at infinite-temperature and then quenched at its
critical temperature $T_c$ at $t=0$. Three lattices sizes were investigated: 
$L=128$, $192$ and $256$ in order to check the possibility of finite-size effects.
Because of the particular structure of their ground-state, the multispin Ising model
and the Baxter-Wu model were simulated on lattices of size $L=258$, a multiple of $3$.
The data have been averaged over $10,000$ initial configurations for $L=128$,
$20,000$ for $L=192$ and $50,000$ for $L=256$ and $L=258$. These parameters
are the same for all simulations presented in this paper. We measured the autocorrelation
functions $C(t,s)$ and the fluctuation-dissipation ratio $X(t,s)$ for $t\le 1500$
and $s=10$, $20$, $40$, $80$ and $160$. For all observables, errors bars on the
averages over the initial configurations have been estimated as the standard deviation.
In principle, the response functions could be used to estimate the exponent
$\lambda/z$ too but they turn out to be smaller and thus noisier than autocorrelation
functions making the numerical estimates of exponents much less accurate.

\begin{center}
\begin{figure}[!ht]
        \centerline{\psfig{figure=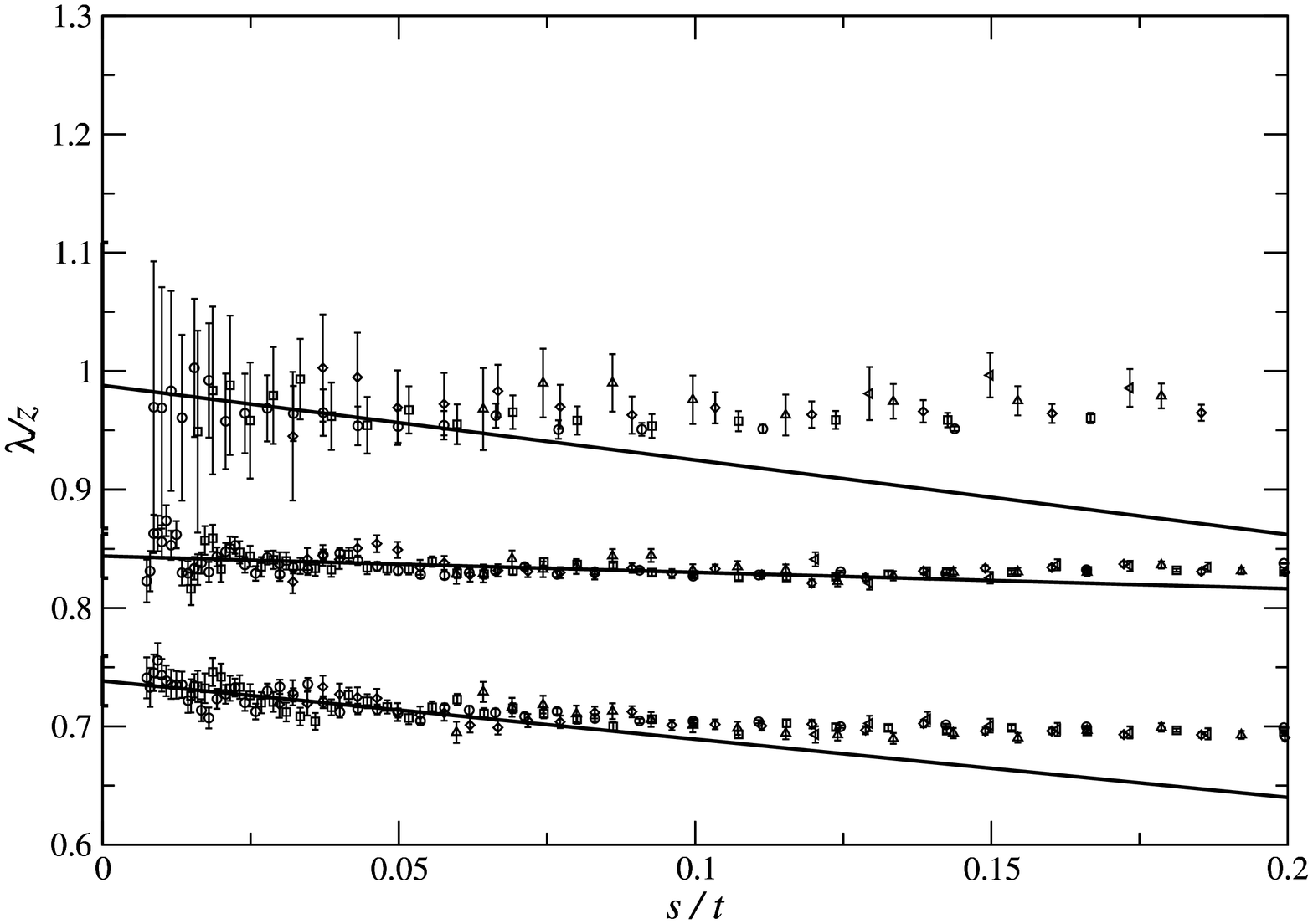,height=6cm}}
	\centerline{\psfig{figure=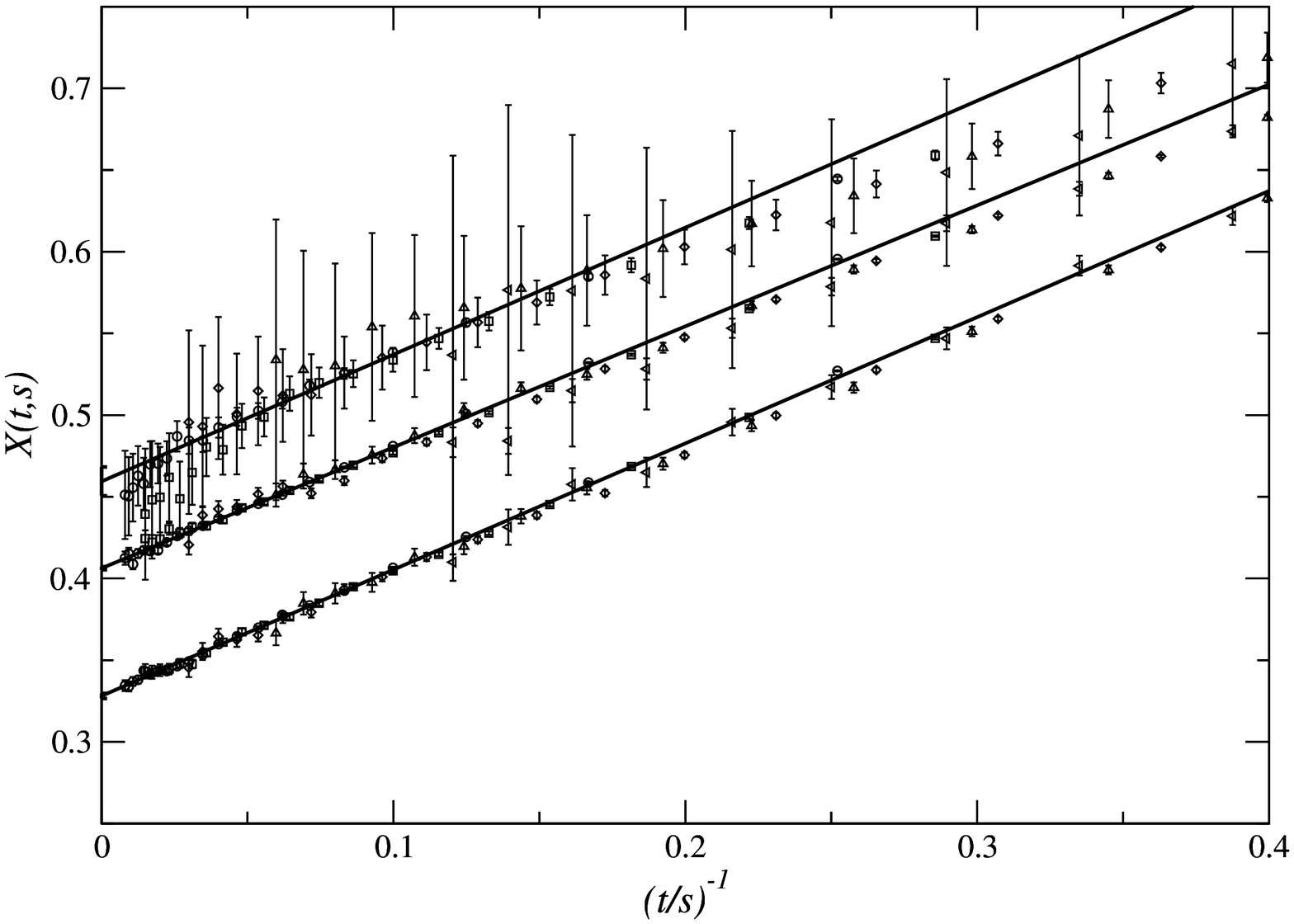,height=6cm}}
        \caption{Top: effective critical exponent $\lambda/z$ calculated by
	power-law interpolation of $C(t,s)$ over the range $t\in[t_{\rm min};1500]$
	versus $s/t_{\rm min}$ for the Ising model, the 3-state clock
	and the 4-state Potts model (from bottom to top). Bottom: fluctuation-dissipation
	ratio $X(t,s)$ versus $s/t$ for the same models (from bottom to top).
	The different symbols correspond to $s=10$ (circle), $s=20$ (square),
	$s=40$ (diamond), $s=80$ (triangle up) and $s=160$ (triangle left). The lattice
	size is $L=256$.}
        \label{fig2}
\end{figure}
\end{center}

Data for the autocorrelation function $C(t,s)$ have been grouped into bins of twenty points (forty
for the 4-state Potts model whose fluctuations are much larger than other models). For each
bin, an effective exponent $\lambda/z$ was measured by power-law interpolation over the points
inside the bin. We took into account error bars by weighting each point with the inverse of
its square error in the fit. The effective exponent can be considered local because values of
$t/s$ remain very close for points inside the same bin. The effective exponent is plotted on
Figure~\ref{fig2} (on the left) versus the inverse mean position $s/t$ of the bin. A fairly good
collapse of curves corresponding to different values of $s$ is observed for small values of $t/s$.
The main difficulty in the determination of the asymptotic behaviour
is that the effective exponent decreases down to a plateau before slowly increasing as $s/t$ is
going to zero. Moreover, autocorrelation functions are getting smaller and thus noisier when
$s/t$ is going to zero, adding to the difficulty. Our final estimate for $\lambda/z$ is 
the intercept given by a linear fit over points in the range $s/t\in[0;0.05]$, corresponding to any
value of $s$ and with a weight corresponding to the inverse square error. The bold line on
Figure~\ref{fig2} corresponds to this fit. The drawback of this method is that the points with the
smallest error bars, i.e. giving the largest contribution to the fit, correspond to the smallest values of $s$.
As a consequence, problems may be caused by corrections to scaling depending on $s$ only. However,
as can be seen on Figure~\ref{fig2}, the good collapse of curves corresponding to different
values of $s$ suggests that these corrections are weak.
The data having been produced by Markov chains, the effective exponents, measured at
different values of $t$ or even $s$, are correlated. As a consequence, the linear fit underestimates
the true error. The assumption of an exponential decay of the autocorrelations
$G\left({t\over s},{t'\over s'}\right)=\left[{\lambda\over z}\big(t/s\big)-b-a{s\over t}\right]
\left[{\lambda\over z}\big(t'/s'\big)-b-a{s'\over t'}\right]$ where $a$ and $b$ are the
parameters given by a linear fit allows for the estimation of the correlation length $\ell$ by summing
the autocorrelations: $\ell=\sum_{t',s',t,s}G\left({t\over s},{t'\over s'}\right)/
\sum_{t,s}G\left({t\over s},{t\over s}\right)$ in the range $s/t\in[0;0.05]$. The standard deviation
on the intercept $b$ as given by the linear fit is valid
only in case of uncorrelated data for which it decays as $1/\sqrt N$ where $N$ is the number
of degrees of freedom of the fit. To take into account the fact that only one point out of $\ell$ is
statistically independent, we multiplied the error on $X_\infty$ by $\sqrt\ell$.
The correction is quite large since in the case of the Ising model for instance,
the error is $3.10^{-3}$ when neglecting correlations and $2.10^{-2}$ when taking them
into account. On Figure~\ref{fig2}, our final estimate of the exponent $\lambda/z$ is given
by the intercept of the bold line with the $y$-axis and its error bars are plotted along the
$y$-axis. The values are collected in Table~\ref{table1}. They are compatible within error bars
with the values found in the literature~: $\lambda/z\simeq 0.731(3)$~\cite{Grassberger95}
or $0.732$~\cite{Okano97} using $z=2.1667(5)$~\cite{Nightingale00} for the Ising model,
$0.828(2)$~\cite{Schuelke95} for the 3-state Potts model and $0.919$ assuming
$z=2.294$~\cite{Arashiro03} for the 4-state Potts model.
\\

The scaling behaviour of $C(t,s)$ and $R(t,s)$, as given by Eq.~\ref{ScalingCR}, 
leads to a fluctuation-dissipation ratio $X(t,s)$ depending only on $t/s$~:
	\begin{equation}
		X(t,s)={k_BT R(t,s)\over{\partial\over\partial s}C(t,s)}
		\sim {f_R\left({t\over s}\right)
		\over {2\beta\over\nu z}f_C\left({t\over s}\right)
		+{t\over s}f'_C\left({t\over s}\right)}.
	\end{equation}
Renormalisation-group calculations for the $O(n)$ model confirm that
hypothesis~\cite{Gambassi02}. Our numerical estimates of $X(t,s)$ indeed collapse
when plotted versus $t/s$. The asymptotic value $X_\infty=\lim_{s\rightarrow +\infty}
\lim_{t\rightarrow +\infty} X(t,s)$ may thus be obtained as $\lim_{t/s\rightarrow +\infty}
X(t,s)$. Note however that some exactly-solvable models display a cross-over when
$t$ becomes large so that the appropriate regime to take into account might not be
$s/t$ going to zero~\cite{Berthier03}. In our case, no significant dependence on the
value of $s$ is observed: the curves corresponding to different values of $s$ can not be
distinguished within statistical fluctuations and the fluctuation-dissipation ratio $X(t,s)$
displays a nice linear behaviour over a large range of times as can be seen on
Figure~\ref{fig2} (on the right). To lighten the figures, averages over twenty points have
been plotted. On the other hand, all data have been used in the fitting procedure.
The asymptotic value $X_\infty$ of the fluctuation-dissipation ratio is obtained
by a linear fit of $X(t,s)$ with respect to $s/t$ in the range $s/t\in [0;0.15]$.
Again, the data being correlated, standard error on the coefficients of the fit
underestimates the true error. Taking into account autocorrelations of the estimates
of $X(t,s)$ leads to an important correction to the standard deviation on coefficients
given by the linear fit~: in the case of the Ising model, the linear fit gives an error on
$X_\infty$ of $4.6.10^{-4}$ while taking into account correlations, it is estimated to be
$1.5.10^{-3}$. The interpolated line is the bold line on Figure~\ref{fig2} and
the error bar of our final estimate of $X_\infty$ has been put along the $y$-axis.
Our final estimates of $X_\infty$ are collected in Table~\ref{table1}. 
In the case of the 4-state Potts model, the fluctuation-dissipation ratio $X(t,s)$ is
very noisy. However, a rather precise estimation of $X_\infty$ is obtained because not all
points have large error bars so that the main contribution to the fit is due to points with
smaller error bars, in practice points corresponding to small values of $s$. Again, the
procedure may be problematic if corrections depending on $s$ are important. This seems
not to be the case at regard of the good collapse of curves corresponding to different values
of $s$. More problematic is the fact that Figure~\ref{fig2} seems to indicate a downward curvature at
small values of $s/t$. The linear interpolation lies outside the error bars for the points
with the smallest values $s/t$.
Reducing the range of the fit leads indeed to smaller and smaller values of $X_\infty$~:
$0.459(8)$ for $s\in[0;0.15]$, $0.455(8)$ for $s\in[0;0.1]$ and $0.36(13)$ for
$s\in[0;0.05]$. Note that equilibrium quantities display logarithmic corrections that may also
be present for out-of-equilibrium ones. On the other hand, the data being correlated, this curvature
may not be a general trend but a fluctuation. Since all this is highly hypothetic, we will
adopt the safer attitude and consider the linear fit inappropriate in this case. 

\begin{table}[!ht]
\begin{center}
\begin{tabular}{@{}*{4}{l}}
Models & $\lambda/z$ & $X_\infty$ \\
\hline
Ising model		& 0.738(21) & 0.328(1)\\
3-state clock model 	& 0.844(19) & 0.406(1)\\
4-state Potts model	& 0.99(12)  & $0.459(8)^\dagger$\\
\end{tabular}\end{center}
\caption{Exponent $\lambda/z$ and fluctuation-dissipation ratio $X_\infty$
for the Ising model, the 3-state clock model and the 4-state Potts
model. $^\dagger$ The value of $X_\infty$ given for the 4-state Potts model has to be
considered carefully since a deviation from a purely linear behaviour is observed.}
\label{table1}
\end{table}

\section{Universality for different lattices}
\begin{center}
\begin{figure}[!ht]
        \centerline{\psfig{figure=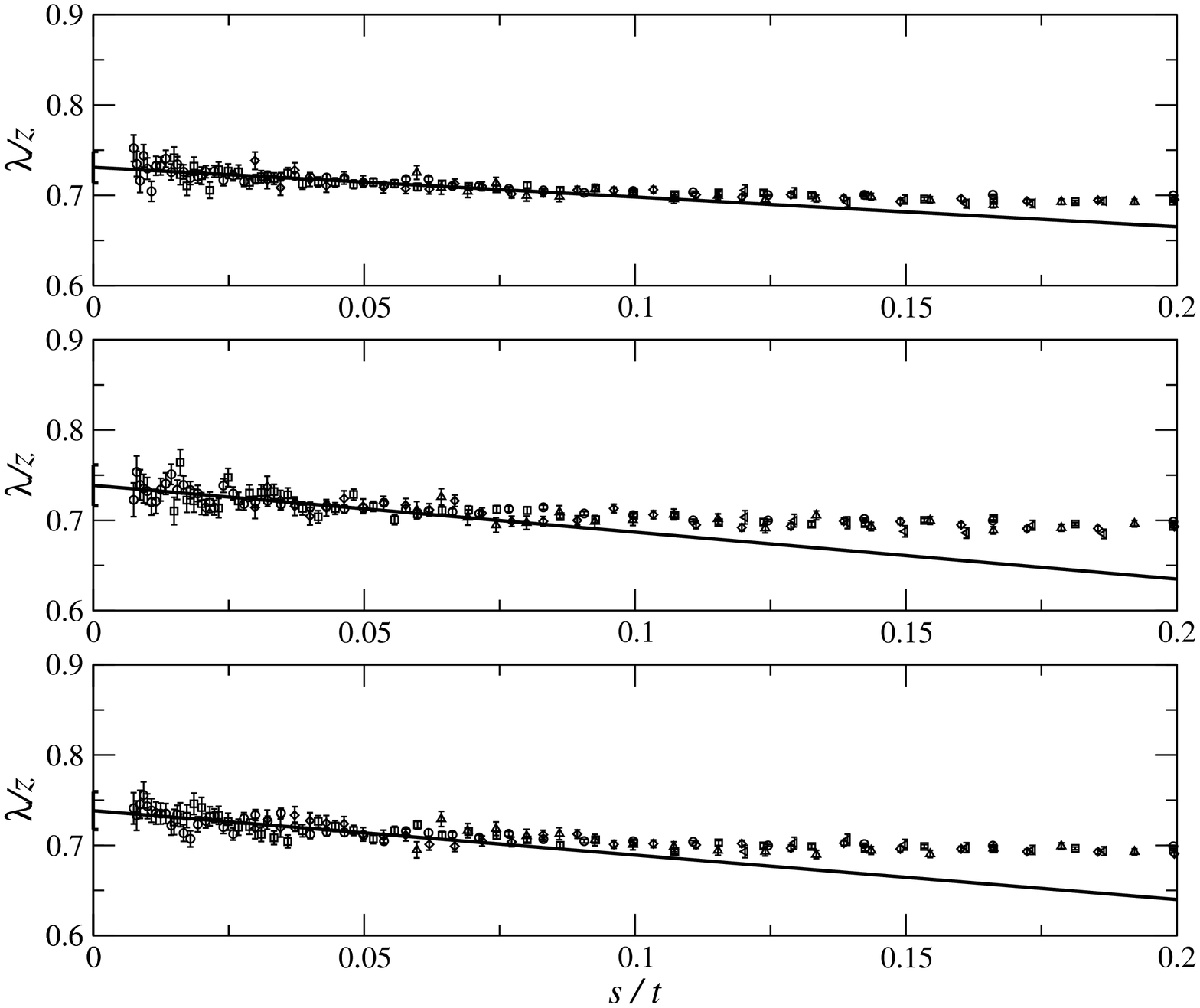,height=6cm}}
	\centerline{\psfig{figure=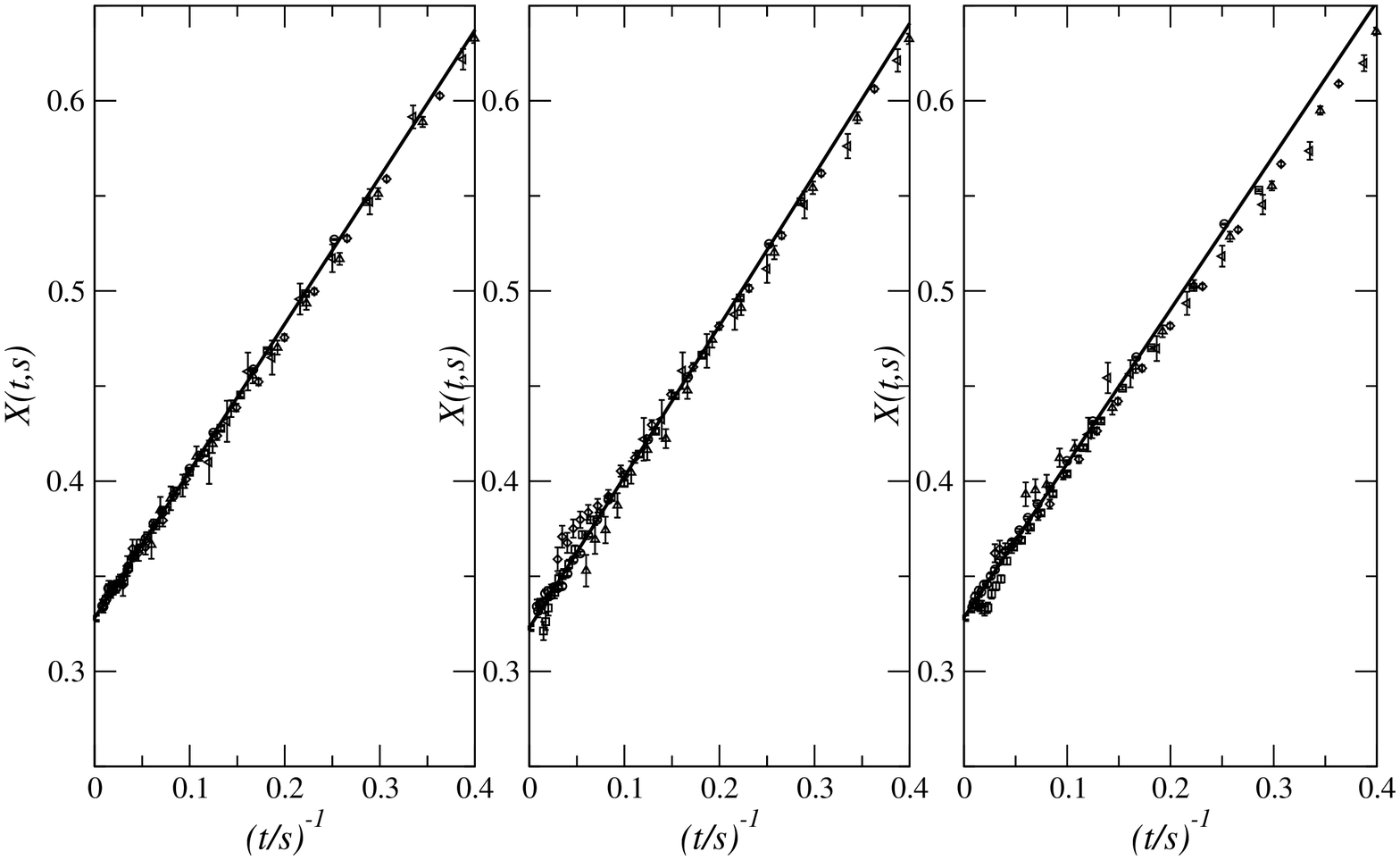,height=6cm}}	
        \caption{Top: effective critical exponent $\lambda/z$ calculated by
	power-law interpolation of $C(t,s)$ over the range $t\in[t_{\rm min};1500]$
	versus $s/t_{\rm min}$ for the Ising model on the square, triangular and
	honeycomb lattices (from bottom to top). Bottom: fluctuation-dissipation
	ratio $X(t,s)$ versus $s/t$ for the same models (from left to right).
	The different symbols correspond to $s=10$ (circle), $s=20$ (square),
	$s=40$ (diamond), $s=80$ (triangle up) and $s=160$ (triangle left). The lattice
	size is $L=256$.}
        \label{fig4}
\end{figure}
\end{center}

\begin{center}
\begin{figure}[!ht]
        \centerline{\psfig{figure=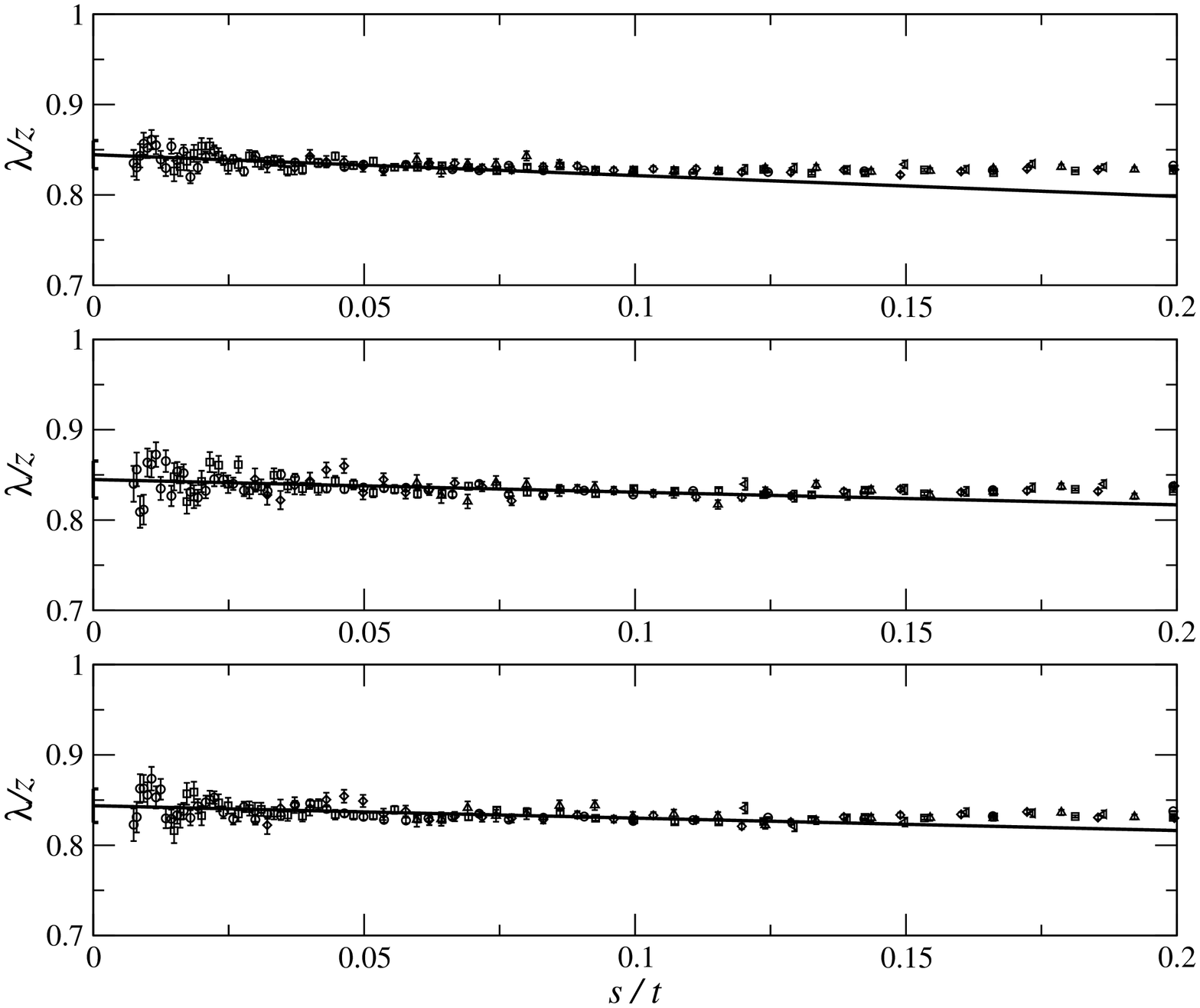,height=6cm}}
	\centerline{\psfig{figure=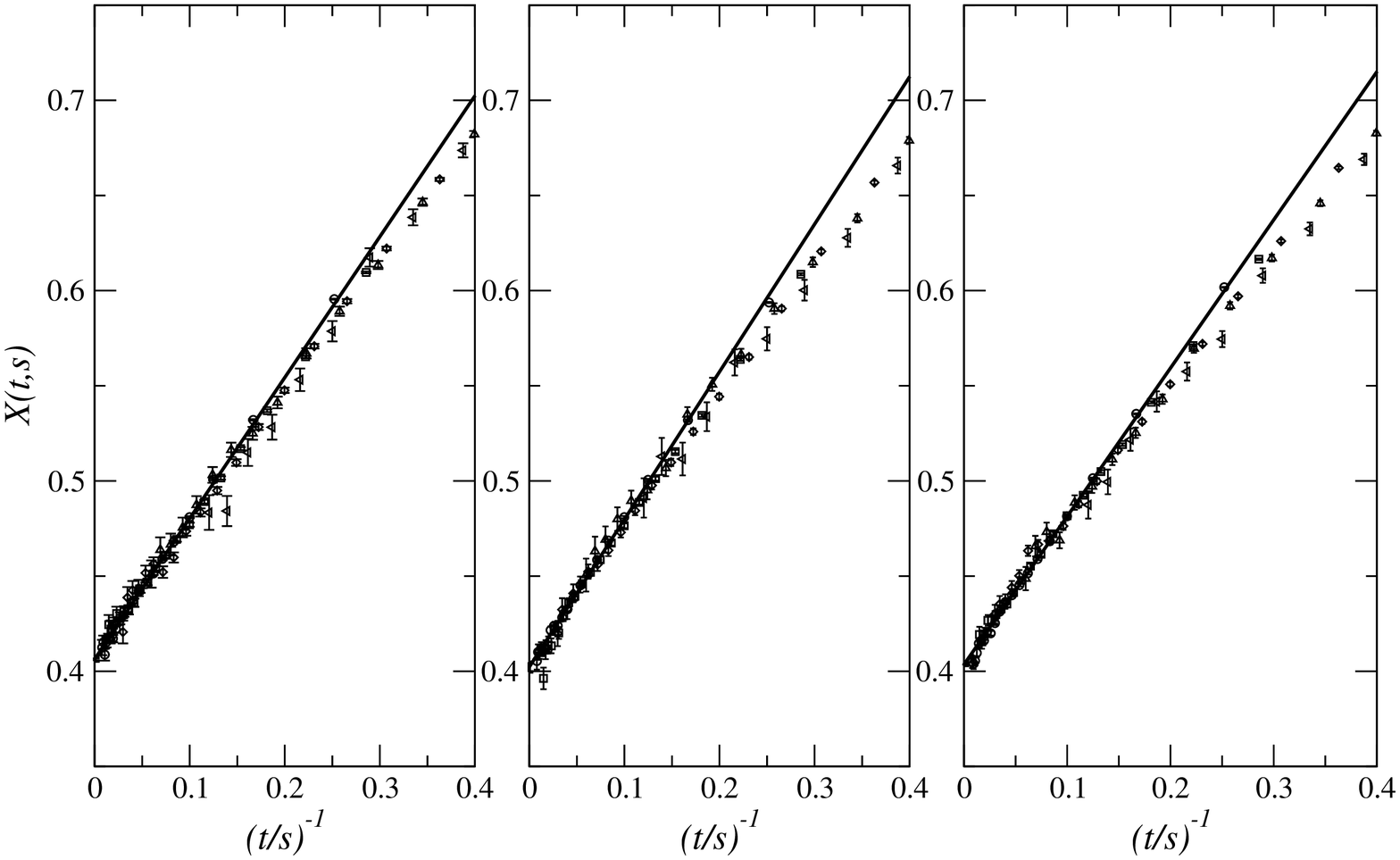,height=6cm}}
        \caption{Top: effective critical exponent $\lambda/z$ calculated by
	power-law interpolation of $C(t,s)$ over the range $t\in[t_{\rm min};1500]$
	versus $s/t_{\rm min}$ for the 3-state clock model on the square, triangular and
	honeycomb lattices (from left to right). Bottom: fluctuation-dissipation ratio
	$X(t,s)$ versus $s/t$ for the same models (from left to right).
	The different symbols correspond to $s=10$ (circle), $s=20$ (square),
	$s=40$ (diamond), $s=80$ (triangle up) and $s=160$ (triangle left). The lattice
	size is $L=256$.}
        \label{fig5}
\end{figure}
\end{center}

\begin{center}
\begin{figure}[!ht]
        \centerline{\psfig{figure=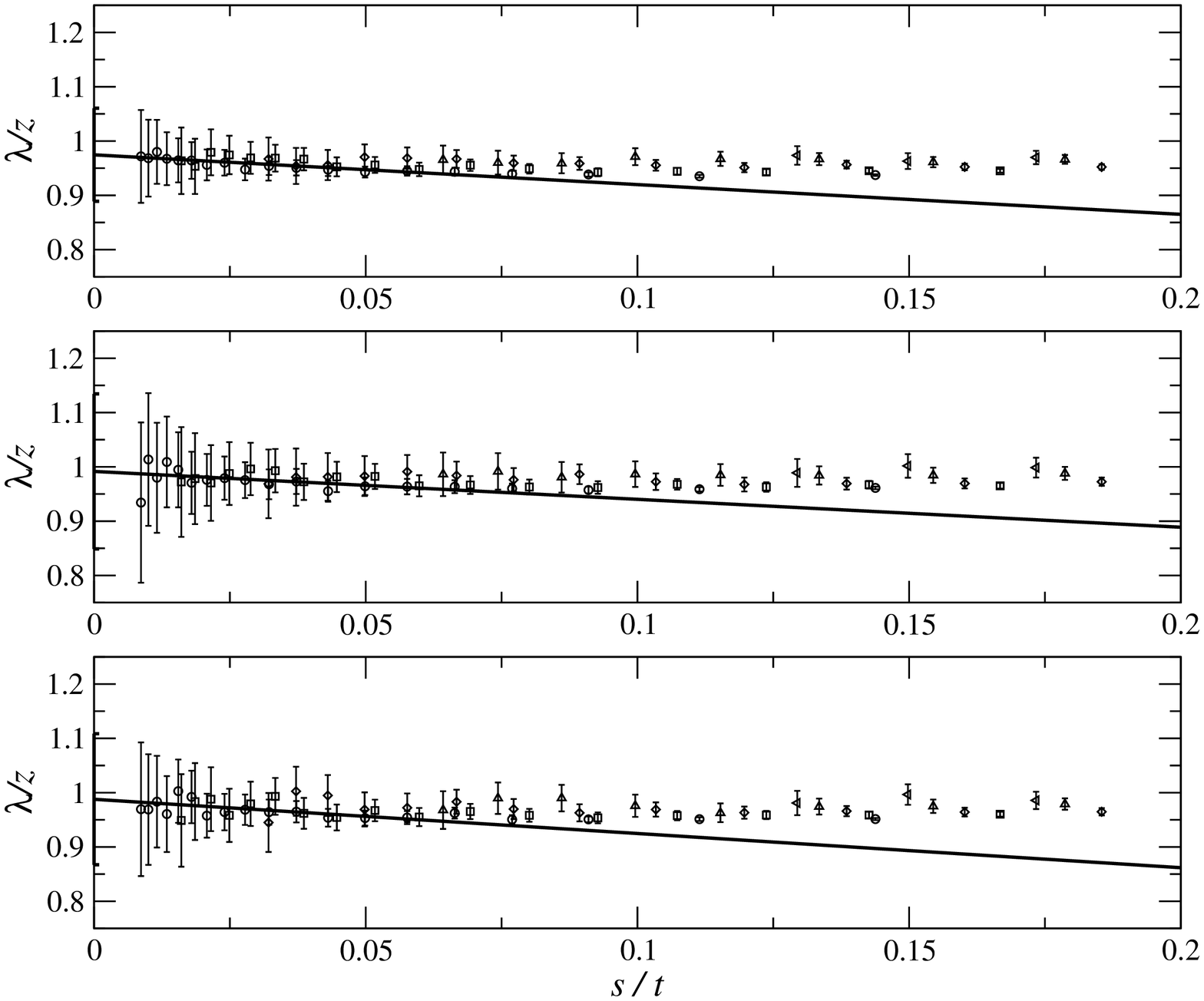,height=6cm}}
	\centerline{\psfig{figure=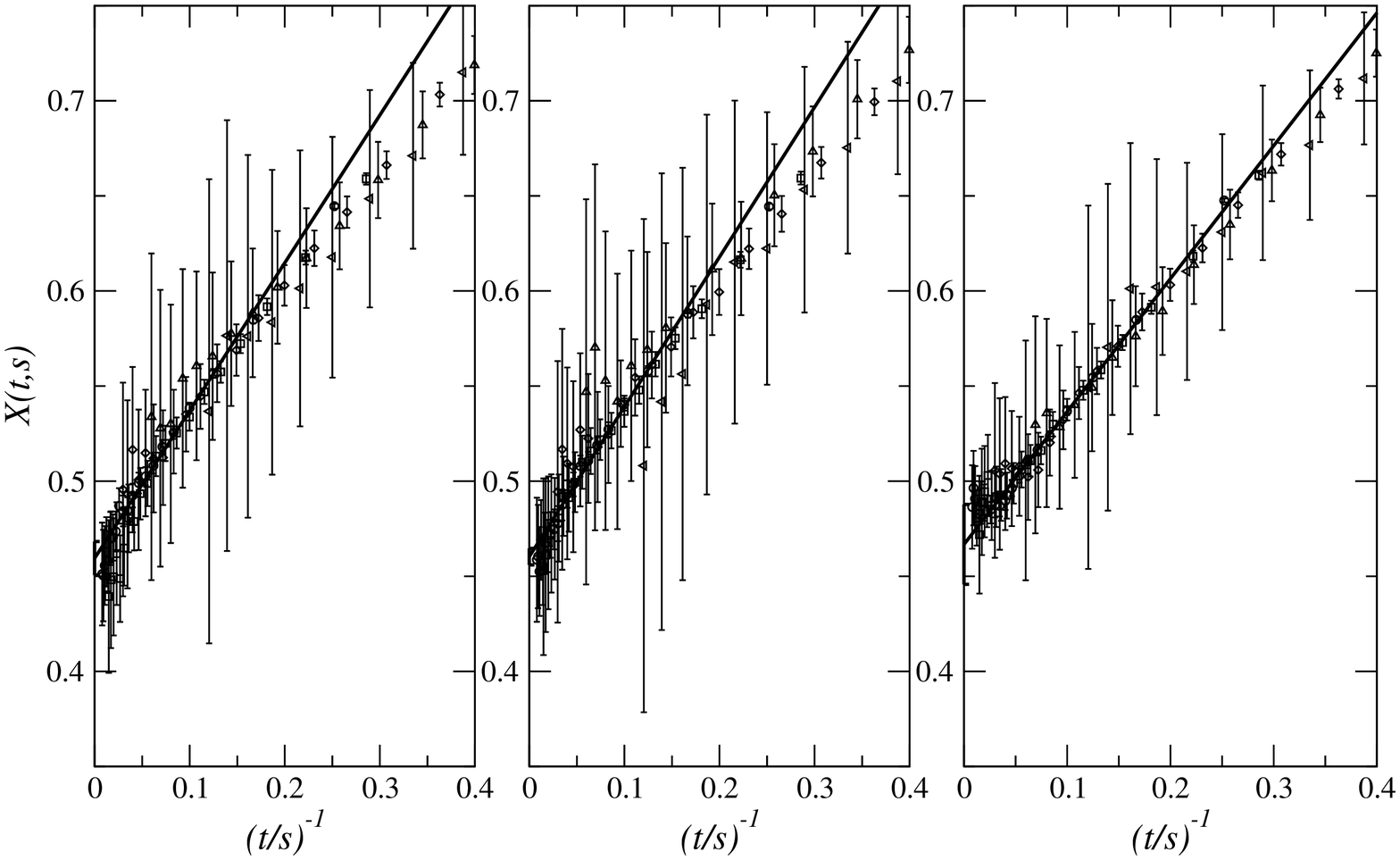,height=6cm}}
        \caption{Top: effective critical exponent $\lambda/z$ calculated by
	power-law interpolation of $C(t,s)$ over the range $t\in[t_{\rm min};1500]$
	versus $s/t_{\rm min}$ for the 4-state Potts model on the square, triangular and
	honeycomb lattices (from left to right). Bottom: fluctuation-dissipation ratio
	$X(t,s)$ versus $s/t$ for the same models (from left to right).
	The different symbols correspond to $s=10$ (circle), $s=20$ (square),
	$s=40$ (diamond), $s=80$ (triangle up) and $s=160$ (triangle left). The lattice
	size is $L=256$.}
        \label{fig6}
\end{figure}
\end{center}

The procedure detailed in the previous section for the Ising model, the 3-state
clock model and the 4-state Potts model on square lattices was extended to triangular
and honeycomb lattices. Note that we only considered regular periodic lattices since
aperiodic or random lattices may change the universality class at equilibrium.
The results are presented on Figure~\ref{fig4} for the Ising
model, Figure~\ref{fig5} for the 3-state clock model and Figure~\ref{fig6} for the 4-state
Potts model. The effective exponents $\lambda/z$ and the fluctuation-dissipation ratios
$X_\infty$ are collected in Table~\ref{table2}. Exponents $\lambda/z$ appear to be
lattice-independent. This statement is in agreement with measurements of the dynamical
exponent $z$ for the Ising model at $T=T_c$ on square, triangular and honeycomb
lattices~\cite{Wang97}. Apart that of the Ising model on triangular lattice, our estimates
of the fluctuation-dissipation ratios for different lattices are compatible within error bars.

The data for the 4-state Potts model are unfortunately too noisy for the estimates of $\lambda/z$
to be really useful. The downward curvature of $X(t,s)$ at small values of $s/t$ observed on
square lattice for this model is also present on triangular lattice but not on honeycomb
lattice where on the other hand a small upward curvature is observed. Despite of these curvatures,
all extrapolated values of $X_\infty$ are compatible within error bars.

\begin{table}[!ht]
\begin{center}
\begin{tabular}{@{}*{7}{l}}
 & & $\lambda/z$ & & & $X_\infty$ & \\
Models & Square & Triang. & Honeyc. & Square & Triang. & Honeyc. \\
\hline
Ising		& 0.738(21) & 0.739(22) & 0.731(17) & 0.328(1) & 0.323(1) & 0.328(1)  \\
3-state clock 	& 0.844(18) & 0.845(20) & 0.844(16) & 0.406(1) & 0.402(3) & 0.404(1) \\
4-state Potts	& 0.99(12) & 0.99(17) & 0.97(8) &
			$0.459(8)^\dagger$ & $0.460(4)^\dagger$ & $0.467(21)^\dagger$ \\
\end{tabular}\end{center}
\caption{Exponent $\lambda/z$ and fluctuation-dissipation ratio $X_\infty$
for the Ising model, the 3-state clock model and the 4-state Potts
model on square, triangular and honeycomb lattices. $^\dagger$ The value of $X_\infty$ given
for the 4-state Potts model have to be
considered carefully since in both cases a deviation from a purely linear behaviour is observed.}
\label{table2}
\end{table}

\section{Universality for different models}
\begin{center}
\begin{figure}[!ht]
        \centerline{\psfig{figure=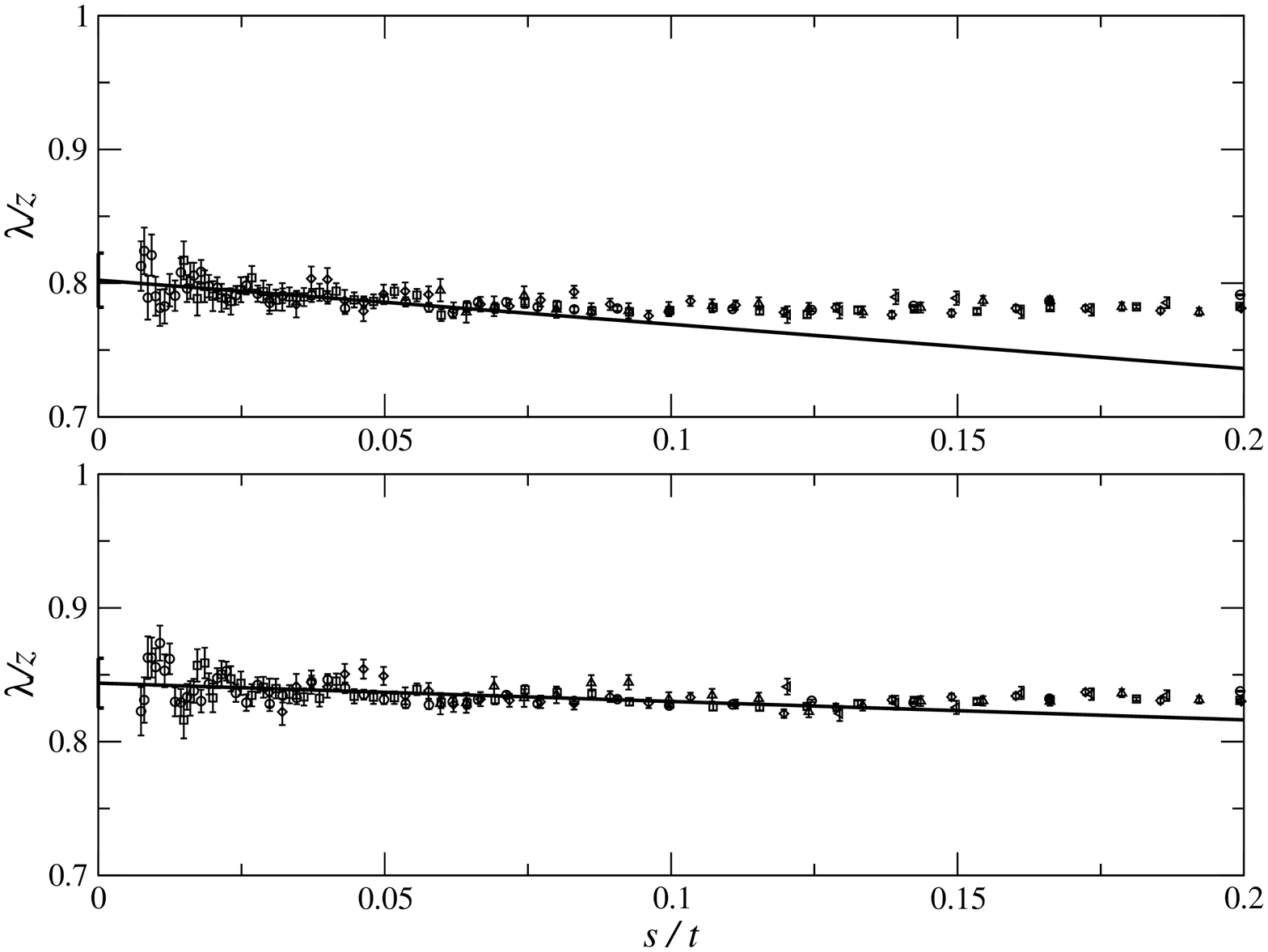,height=6cm}}
	\centerline{\psfig{figure=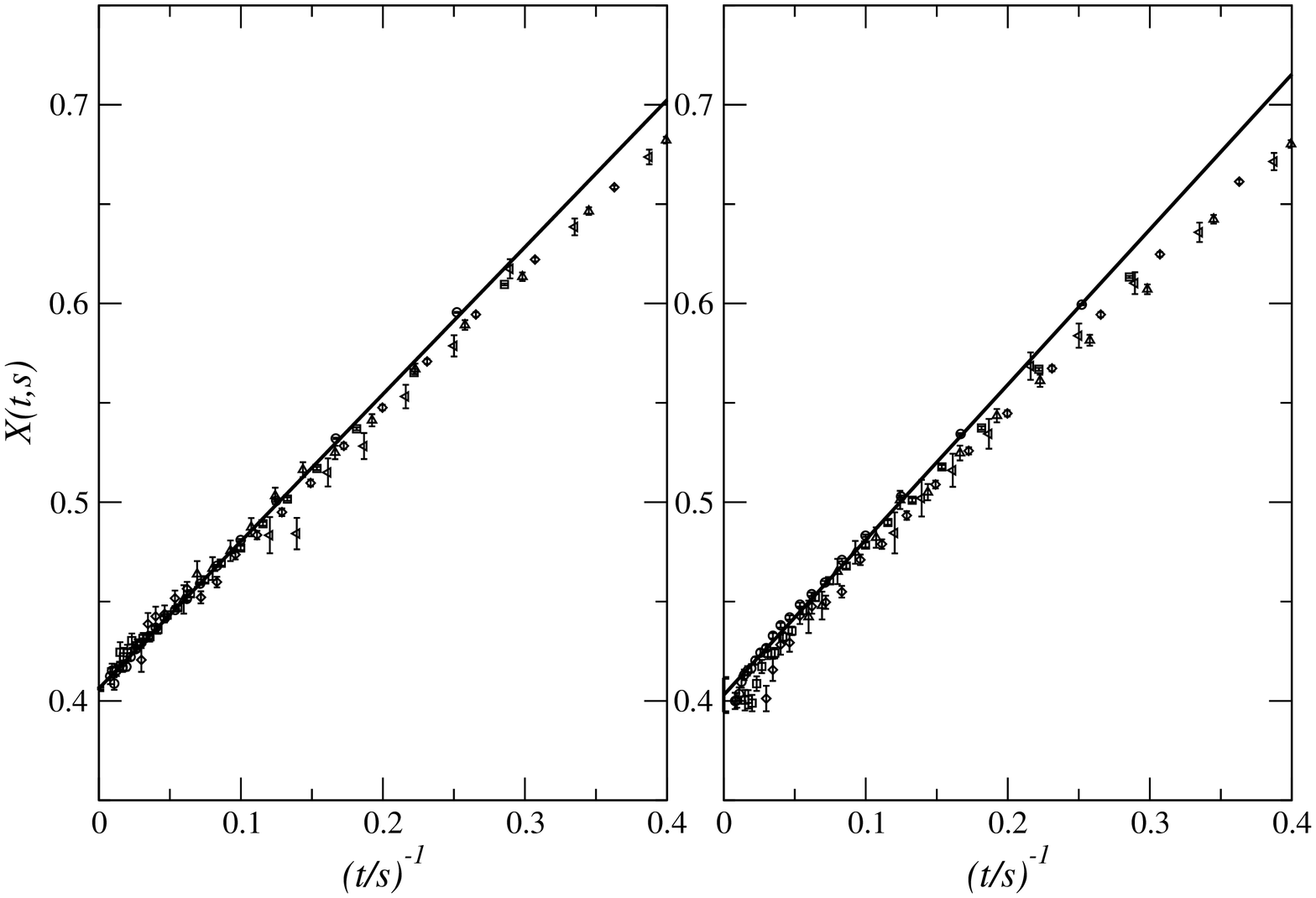,height=6cm}}
        \caption{Top: effective critical exponent $\lambda/z$ calculated by
	power-law interpolation of $C(t,s)$ over the range $t\in[t_{\rm min};1500]$
	versus $s/t_{\rm min}$ for the 3-state clock model and Ashkin-Teller at the self-dual
	point $y=3/4$ (from bottom to top). Bottom: fluctuation-dissipation ratio
	$X(t,s)$ versus $s/t$ for the same models (from left to right).
	The different symbols correspond to $s=10$ (circle), $s=20$ (square),
	$s=40$ (diamond), $s=80$ (triangle up) and $s=160$ (triangle left). The lattice
	size is $L=256$.}
        \label{fig7}
\end{figure}
\end{center}

The procedure was also applied to different models belonging to the same universality class
at equilibrium. We first studied the Ashkin-Teller model at the point of its exactly-known
critical line where the model belongs to the 3-state Potts model universality class. The
effective exponent $\lambda/z$ and the fluctuation-dissipation ratio $X_\infty$ that
we numerically obtained are plotted on Figure~\ref{fig7} and our final estimates are collected
in Table~\ref{table3}.  Our estimates of the exponent $\lambda/z$ are incompatible for the
3-state clock model and the Ashkin-Teller model. Both are compatible within error bars
with the estimates found in the literature~: $0.828(2)$ for the 3-state clock model~\cite{Schuelke95}
and $0.798$ for the Ashkin-Teller when fitting the data given by~\cite{Li97} along the critical line
and assuming a dynamical exponent equal to that of the 3-state Potts model~\cite{Schuelke95}.
While these models belong to the same universality class at equilibrium, it seems thus not to be
anymore the case out-of-equilibrium. Surprisingly, the fluctuation-dissipation ratios $X_\infty$
are compatible within error bars for the two models, although a small downward curvature may be
observed for the Ashkin-Teller model.

\begin{table}[!ht]
\begin{center}
\begin{tabular}{@{}*{4}{l}}
Models & $\lambda/z$ & $X_\infty$ \\
\hline
3-state clock model 	& 0.844(18) & 0.406(1)\\
Ashkin-Teller model 	& 0.802(20) & 0.403(8)\\
\hline
4-state Potts model	& 0.99(12)  & $0.459(8)^\dagger$\\
Baxter-Wu model		& 1.13(6)   & $0.548(15)^\dagger$\\
Multispin Ising model	& 0.977(25)  & 0.466(3)\\
\end{tabular}\end{center}
\caption{Exponent $\lambda/z$ and fluctuation-dissipation ratio $X_\infty$
for different models belonging either to the 3-state Potts model at equilibrium
or the 4-state Potts model.  $^\dagger$ The value of $X_\infty$ given for the 4-state Potts
model and the Baxter-Wu model have to be considered carefully since in both cases a deviation
from a purely linear behaviour is observed.}
\label{table3}
\end{table}

\begin{center}
\begin{figure}[!ht]
        \centerline{\psfig{figure=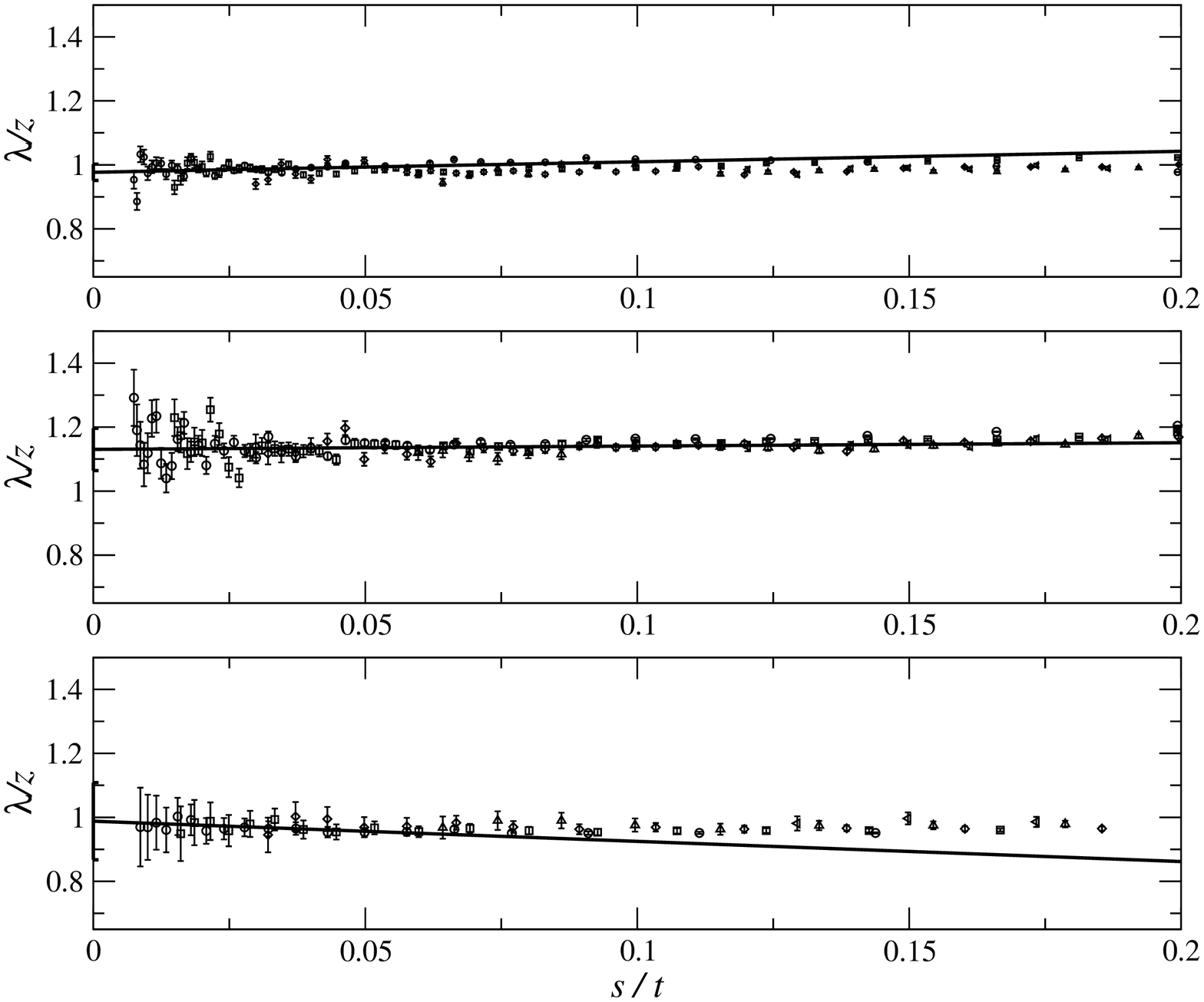,height=6cm}}
	\centerline{\psfig{figure=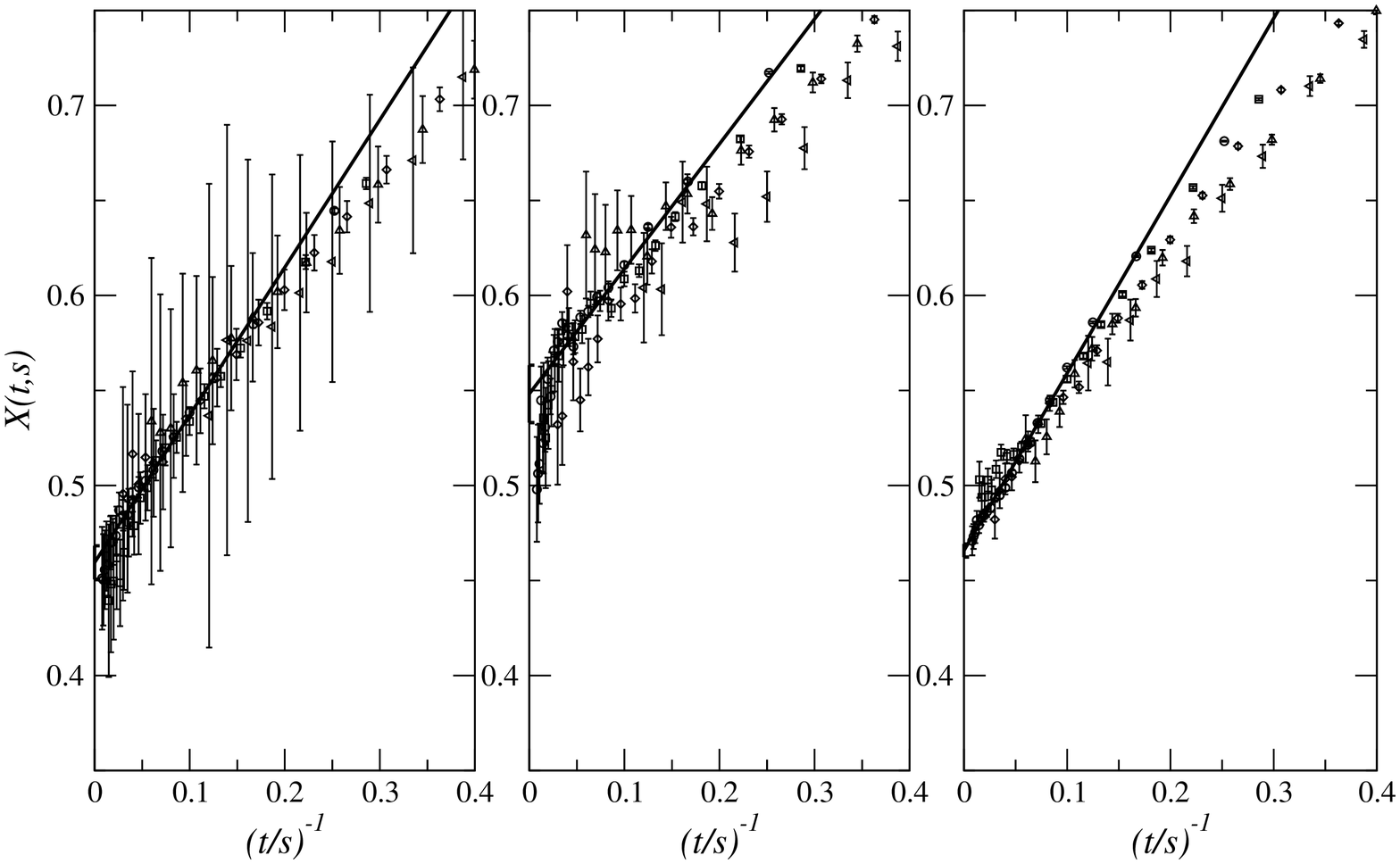,height=6cm}}	
        \caption{Top: effective critical exponent $\lambda/z$ calculated by
	power-law interpolation of $C(t,s)$ over the range $t\in[t_{\rm min};1500]$
	versus $s/t_{\rm min}$ for the 4-state Potts model, the	Baxter-Wu model and a
	multispin Ising model (from bottom to top). Bottom: fluctuation-dissipation
	ratio $X(t,s)$ versus $s/t$ for the same models (from left to right).
	The different symbols correspond to $s=10$ (circle), $s=20$ (square),
	$s=40$ (diamond), $s=80$ (triangle up) and $s=160$ (triangle left). The lattice
	size is $L=256$ for the 4-state Potts model and $L=258$ for the others.}
        \label{fig8}
\end{figure}
\end{center}

We then turned to the study of a multispin Ising model and of the Baxter-Wu model both
belonging to the 4-state Potts model universality class at equilibrium. The
effective exponent $\lambda/z$ and the fluctuation-dissipation ratio $X_\infty$ 
that we numerically obtained are plotted on Figure~\ref{fig8} and our final estimates
are collected in Table~\ref{table3}. Our estimates of $\lambda/z$ for the 4-state Potts model
and the Baxter-Wu model are compatible within error bars but the latter are very large.
On the other hand, estimates of $\lambda/z$ for the multispin
Ising model and the the Baxter-Wu model are fully incompatible. Our values are compatible,
although at the boundary of error bars, with the estimates found in the literature for the
4-state Potts model~\cite{Arashiro03}~: $0.919$ assuming $z=2.294$ and the Baxter-Wu
model~\cite{Arashiro03}~: $1.058(4)$ but not for the multispin Ising model~\cite{Simoes01}~:
$0.902(10)$ assuming the same value for $z$ ($0.870(11)$ with $z=2.380(4)$).
Note that these values were calculated from the estimates of $\theta$ and $z$
obtained by short-time dynamics Monte Carlo simulations. The value of $\lambda/z$
that we give is very sensitive to the accuracy of the dynamical exponent $z$.
Note as well that we would have obtained smaller estimates of $\lambda/z$ if we would have
done shorter simulations or restricted our calculations to smaller values of $t/s$.
The fluctuation-dissipation ratio $X_\infty$ reproduces the same tendency as can be
seen on Table~\ref{table3}. The estimates for the 4-state Potts model and the multispin
Ising model are in agreement while that of the Baxter-Wu model is clearly not. However, the data
for both the 4-state Potts model and the Baxter-Wu model show a downward curvature as can be
seen on Figure~\ref{fig8}. The final estimates of $X_\infty$ have to be taken carefully for
these models.

\section{Conclusions}
We have addressed the question of universality for ageing ferromagnets,
focusing on the study of the autocorrelation decay exponent $\lambda/z$ and the
fluctuation-dissipation ratio $X_\infty$. It turns out that, apart from $X_\infty$
for the Ising model on triangular lattice, both $\lambda/z$ and $X_\infty$ do not
depend on the particular lattice on which the models lives. This supports
the idea of an extension of equilibrium universality classes to out-of-equilibrium
processes. On the other hand, the Baxter-Wu model and a multispin Ising model, both belonging
to the 4-state Potts model universality class at equilibrium, do not share the same
exponent $\lambda/z$. Moreover, the 3-state Potts model and the Ashkin-Teller
model, both belonging to the same universality class at equilibrium, have
different exponents $\lambda/z$ but the same fluctuation-dissipation ratio
$X_\infty$. This raises the question of the relevant quantities sufficient to
characterise the universality class. Let us recall that we have not studied the
influence of the dynamics itself on the universal quantities. It would be
interesting for example to check whether the transition rates Eq. \ref{eq3} and
\ref{eq3b} lead to the same exponents and fluctuation-dissipation ratios or not.
\\

\section*{Acknowledgements}
The laboratoire de Physique des Mat\'eriaux is Unit\'e Mixte de Recherche
CNRS number 7556. The numerical calculations were made in the computing center
CINES in Montpellier under project number pmn2425. The author wishes to express
his gratitude to the members of the statistical group at the university Nancy 1
for useful discussions and to Jean-Marc Luck and Claude Godr\`eche for critical
reading of the manuscript.

\end{document}